# Is the Basic Unit System a String?

**Edgar Paternina.**

Electrical engineer Author of Physics and The Principle of Synergy, published in CD ROM in English and Spanish at Amazon.com.

Contact: epaterni@epm.net.co

 Abstract

The main aim of this paper is to present an overview of the need of a new way of coping the fundamental equations of physics, see the references. Our main aim is to build a new metrics in which both time and space are included, but in some sort of minkowskian union of the two and in such a way that that union preserves an independent reality, and where energy, can be taken as the real fundamental issue of that metrics, instead of the particle concept. A departure from General Theory of Relativity is clear, but then the BUS concept behaves itself as a fundamental string.

Introduction

We are not interested from the beginning in the motion of a particular point in space. Our main aim is to build a new metrics in which both time and space are included, but in some sort of minkowskian union of the two and in such a way that that union preserves an independent reality, and where energy can be taken as the real fundamental issue of that metrics.

The particle point of view in physics is true but partial. And it is this partialness the one we are trying to avoid from the beginning and as so *gravitation* will not occupy *an exceptional position with regard to the other forces, particularly the electromagnetic forces*, but energy as a conceptual starting point will be the one that will occupy an exceptional epistemological position in our constructs.

Being our starting point an integral one we will relate that problem with the need to frame in a unified conceptual scheme  the radical duality of the Universe expressed, in general, in the following dyads:

- the particle and the wave problem expressed dramatically in physics as the momentum and position electron problem of Quantum Mechanics.

- the relation between the whole or form and the part, the so-called generalization-specialization problem

- the qualitative aspects of reality and its quantitative ones

- the relation between rotational movement and linear movement being the latter a special case of the former

- the relation between time and space and the fact they are always in a unified framework or sphere of reality, the so-called space time continuum.

- The relation between two poles in a magnetic field and the fact each one of them cannot be isolated makes the magnetic field a very fundamental one where oneness, openness and wholeness are main features.

- And finally the relation of the signifier and the signified in case of the linguistic sign and as an example that transcends physics but that has the same nature or dynamic pattern.

**A Symbol for differentiating**

These two different orders of reality must be clearly differentiated, that is, separated, so to speak, in different boxes, just as apples and oranges. And for this we need a special symbol of differentiation, but also of another one that integrates them both in a unified framework, having then a Basic Unit System in which both components are integrated. This Bus concept is then a whole/part entity or a holon as defined by Ken Wilber[6], and as so with it a new holonic metrics emerges. The symbol for separation was discovered by Cardano in 1545, when trying to solve the simple algebraic equation

$x^2 + 1 = 0$

and the symbol for its integration or the interdependence of the two state variables was found by Leonard Euler in 1745 when studying infinite series.

$$e^{J(\theta)} = \cos(\theta) + J*\sin(\theta) \quad (1)$$

Apparently the cos and sin functions seem to be the same mathematical functions, except by the fact they differ by 90 degrees. But they do indeed are different from each other not only by that fact -which is some way to escape from unidimensionality- but also by the way they change their sign by changing their angle θ, so

Cos(θ) = Cos(-θ)

Sin(θ) = -Sin(θ)

having this second one two solutions, the plus and minus sign of the square root of minus one problem. We can say then that the first component of ER is nondual, and the second one is dual. The first one has to do with the whole, the second one with the part or a binary logic. The first one with a geometric or just a graphical representation and the second one with its corresponding algebraic or mathematical representation. The first one with the dynamic nature or time and the second one with space.

From the physical point of view we need a metrics where we have two state variables, the well-known S and T variables, one related with space, and the other one, the dynamic part of reality, related with time. We will replace them though by E and R, because we will use S for representing the whole physical reality or what Minkowski named the quadratic differential expression, which is an invariant or else it represents the same nature of reality, physical reality in this case. In this sense reality is independent from the observer or its frame of references and *space by itself, and time by itself, are doomed to fade away into mere shadows* as Minkowski put it in his classical paper Space and Time[7]. We have then a fifth sphere of reality as an intermediate domain between the observer and the object that we have named the sphere of Form, where reality can be represented not as something relative but as a new metrics in which the laws of nature hold good independently of the system of reference as the holon concept is a Basic Unit System that preserves the isomorphic properties of all systems.

**A New Sphere of Reality and the Complex Plane**

The introduction of this sphere of reality is necessary both from the point of view of Euler Relation where we have the complex plane by just assigning θ a value of 90 degrees, and from the point of view of the need of a metrics with which we can represent the space time continuum so we will have a new differential quadratic element defined as:

$$DS = Abs(DS) * e^{J(\theta)} \quad (2)$$

that can be represented in rectangular coordinates as

$$DS = dE + J * dR \quad (2)$$

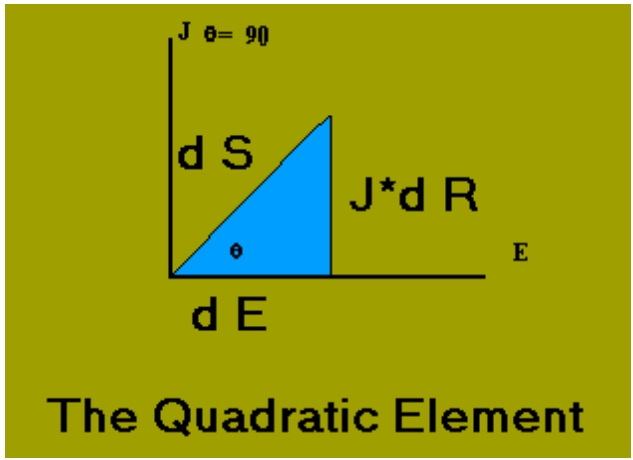

The Quadratic Element

that gives us according to the Pythagorean Theorem

$$DS^2 = dE^2 - dR^2 \quad (3)$$

where $dR^2 = dx^2 + dy^2 + dz^2$

and it is this integral space time mathematical representation the one that permits us to find again all the fundamental equations of physics, including the well-known Schrödinger Wave Equation[1,2,3,4,5]. With this new concept of unit in which the part and the whole are included, we have defined a new metrics with which reality can be represented without the reductionistic drawback, because the Uncertainty Principle is included. But on the other side the concept of dimension acquires its real connotation associated with the four dimensional space time continuum, where for having closed system and objects we need to solve (2) or (3), that is, we need to find laws or relations between the two main component of the BUS concept. And for this we need to consider not only the observed manifestation of that entity represented by the BUS, associated in some cases with a clear graphical representation, but also with the chance it has a corresponding and almost exact mathematical representation, such as in a planet, where on the one side we have an ellipse and on the other we have a corresponding equation for it. In this case we have a closed system with its state completely determined, and when this is the case we can make predictions, we can make measurements.

The main departure of this new approach in regard to the well-known approach due to Einstein where (3) or the "linear element" was generalized as

$$DS^2 = \Sigma G_{ij} dX_i * dX_j$$

is precisely to reduce the whole problem to find 10 functions $G_{ij}$ according to the rules of Tensor Analysis, so to speak, to just an analytical problem, not a geometric problem anymore, as geometry was then reduced to algebra. So Einstein wrote "Thus, according to the general theory of relativity, gravitation occupies an exceptional position with regard to other forces, particularly the electromagnetic forces, since the ten functions represent the gravitational field at the same time define the metrical properties of the space measured.." If we assimilate each $G_{ij}$ to a dimension, mathematical dimension, we will need to posit 10 dimensions to solve the whole problem, and in spite those additional dimensions have no physical meaning at all.

Expression (2) is an infinitesimal rotating vector or pointer or else a tiny loop or string vibrating at certain frequency, but it can also vibrate at other modes or frequencies, but in the complex plane. The fact that with it we can deduce the Schrödinger Wave Equation takes us to think that the main and fundamental concept behind everything physical is energy, and defined like a frequency multiplied by Planck's constant h, and not precisely the particle concept, even though the latter is a derived one, and an electron can be seen just as a whirlpool of energy.

The fact that with (2) and (3) we can find all the fundamental equations of physics as is shown in the references, makes it possible for us to have ways to express the same general laws of nature or dynamic patterns from a unified conceptual point of view, which was the central claim of the systems sciences, so the unity of science is granted not by the reduction of all sciences to physics but by those isomorphic regularities[5] of the different levels of reality, including the physiosphere or the so-called four dimensional space time continuum and the biosphere too as in that new domain we have named the domain of form and represented by the complex plane, life can be defined as an animated form.

The problem of two arrows of time and the Second Law of Thermodynamics is associated with the definition of open systems, and in this case that problem is overcome because with (2) we have found the equations of the pendulum, a truly open system.

**The Principle of Synergy**

The production of electrical energy or Alternating Current is a real example of what I have called the application of the principle of synergy. In that production we have a threefold physical magnetic structure rotating because of the hydraulic turbines and at the end, what we have is a rotating magnetic field that by Maxwell Laws induces in each one of the terminals of that threefold structure the three phase AC we use daily. Mathematically we can represent this situation by considering three dyads of space and time vectors, that is,

$$A = Abs(A) * e^{J(\theta)} + Abs(A) * e^{J(wt)}$$

$$A = Abs(A) * e^{J(\theta+120)} + Abs(A) * e^{J(wt)}$$

$$A = Abs(A) * e^{J(\theta+240)} + Abs(A) * e^{J(wt)}$$

by using ER we decompose each one of the above equations in sin and cos, and after summing them up we finally obtain that the three space components become null, so we have just a dynamic expression:

$$A = Abs(A') * e^{J(wt)}$$

By mathematical inevitability we can represent that sum as in the figure

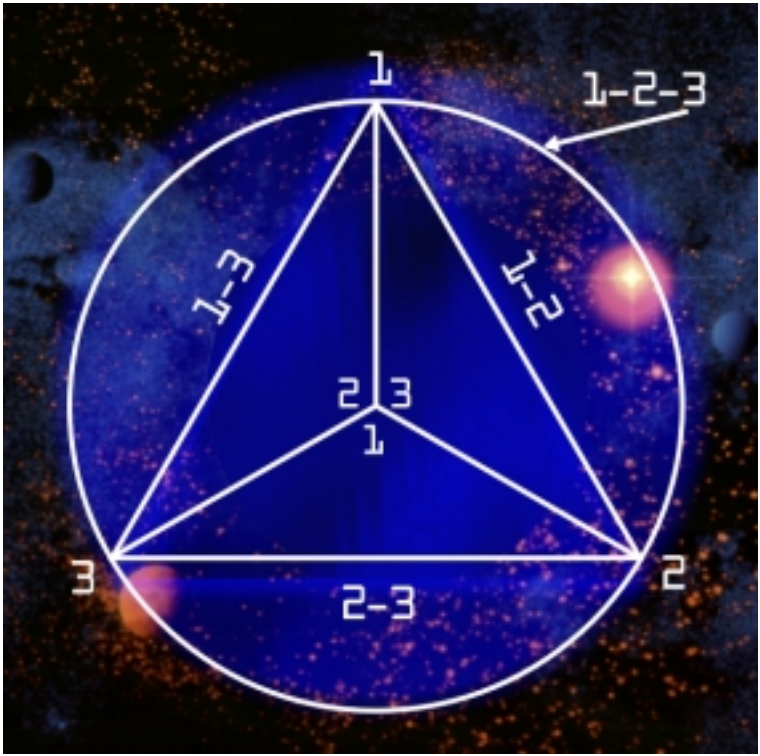

What we have here is just a sum 1-2-3 that is greater than the sum of its parts and what this means is that we have an open system that is interchanging energy with the environment, so in this sense this is something practical, as the extra that can be in the whole that is not in the parts comes from the environment, from the field concept as a rotating entity and which is a complete whole by itself just in case of the magnetic field, a reason why it seems so fundamental in the universe.

Synergy is today a fashionable catchword, a word we hear almost everyday in the business environment and it means precisely that sum that is greater than the sum of its parts. In this sense it is synergy with the help of the field concept and the Basic Unit System or Holon concept, the one that permits us to define

Entropy = f (-Synergy)

so that problem of the two time arrows is not our problem anymore, and Entropy is just the result of the not application of that principle of synergy. If we apply this principle we can have emergent states, or movements as that of the pendulum that seemed to violate the infamous Second Law of

Thermodynamics. Before finishing we want to point out some of the main features of this new metrics

- We do not need to use Tensor Analysis whose main disadvantage being its abstraction from the point of view of physical representation
- We do not need then to introduce additional strange dimensions, with no physical meaning at all
- A Basic Unit System by definition is a rotating entity but in the complex plane and as such it behaves like a string
- The complex representation of the BUS concept reduces complexity by minus one degree[5]
- We have with the BUS concept a new way of presenting the Uncertainty Principle [1,5]
- We have a new way of finding the fundamental equations of physics in the line of reasoning of that claim put by the systems sciences
- We have then a new holonic worldview by introducing an intermediate level of reality between the observer and the object that definitely permits us to exorcise the ghost of consciousness from physics.

References

1. Epsilon Pi. **Physics and The Principle of Synergy** Amazon.com.1999
2. **The Basic Unit System Concept and the Principle of Synergy** ( http://xxx.lanl.gov/html/physics/9908040 )
3. **Electromagnetism, Relativity and the Basic Unit System**
( http://xxx.lanl.gov/html/physics/9908042)
4. **Gravitational Fields and The Basic Unit System concept**
( http://xxx.lanl.gov/html/physics/9908045 )
5. **The Principe of Synergy and Isomorphic Units**
( http://xxx.lanl.gov/html/physics/0010022 )
6. Ken Wilber**. Sex, Ecology and Spirituality. Shambhala Boston & London. 1995**
**7.** Einstein at all**. The Principle of Relativity. Dover Publications, INC.1952**